# Does the truth come naturally? Time pressure increases honesty in one-shot deception games

Valerio Capraro

Middlesex University

London

United Kingdom

V.Capraro@mdx.ac.uk

**Abstract**

Many situations require people to act quickly and are characterized by asymmetric information. Since asymmetric information makes people tempted to misreport their private information for their own benefit, it is of primary importance to understand whether time pressure affects honest behavior. A theory of social heuristics (the Social Heuristics Hypothesis, SHH), predicts that, in case of one-shot interactions, such an effect exists and it is positive. The SHH proposes that when people have no time to evaluate all available alternatives, they tend to rely on heuristics, choices that are optimal in everyday, repeated interactions and that have been internalized over time; and then, after deliberation, people shift their behavior towards the one that is optimal in the given interaction. Thus, the SHH predicts that time pressure increases honesty in one-shot interactions (because honesty may be optimal in repeated interactions, while dishonesty is always optimal in the short run). However, to the best of our knowledge, no experimental studies have tested this prediction. Here, I report a large (N=1,013) study aimed at filling this gap. In this study, participants were given a private information and were asked to report it within 5 seconds vs after 30 seconds. The interaction was one-shot, and payoffs were such that subjects had an incentive to lie. As predicted by the SHH, I find that time pressure *increases* honest behavior. In doing so, these results provide new insights on the role of time pressure on honesty, and provide one more piece of evidence in support of the Social Heuristics Hypothesis.

*Keywords:* lying aversion, honesty, time pressure, deception game.

Introduction

Lying is pervasive in human societies and has enormous undesired economic consequences. For example, tax evasion costs about $100 million to the U.S. government every year (Gravelle, 2009), and, according to the FBI, insurance fraud costs more than $40 million to insurance companies every year[1].

The fact that some people lie when that is beneficial to themselves is not surprising: the standard theory of Homo Economicus assumes that no negative outcomes are associated with the act of lying and thus it explicitly predicts that people would lie, whenever telling a lie would increase their material payoff. However, in contrast to the theory of Homo Economicus, previous research has shown that some people do act honestly and they do so even when lying would be beneficial to all parties involved (Erat & Gneezy, 2012; Cappelen, Sørensen &Tungodded, 2013; Biziou-van-Pol, Haenen, Novaro, Occhipinti-Liberman & Capraro, 2015). These results are particularly interesting, because, in their setting, lying would not only maximize the liar's payoff, but it would also maximize social welfare and minimize inequity. Thus, not only the theory of Homo Economicus predicts that subjects would lie, but also theories assuming that subjects have social preferences for minimizing economic inequalities (Fehr & Schmidt, 1999; Bolton & Ockenfels, 2000) or for maximizing social welfare (Charness & Rabin, 2002; Capraro, 2013) do so. For this reason, these results have been taken as compelling evidence for the fact that individuals have an intrinsic cost of lying. Of course, this cost may be zero for a proportion of "consequentialist" subjects, who, in their decision process, weighs only the economic consequences of their actions and not the actions themselves; but, importantly, the aforementioned findings demonstrate the existence of subjects for whom the cost of lying is not zero: these subjects would lie only if the consequences of deception were "good enough" and, in principle, some of them may even never lie, if they have an infinite cost of lying (deontological subjects). Indeed a rich stream of research has found that the decision to lie in a given context depends on the decision context itself as well as on the decision maker's personal characteristics (Barnes, Schaubroek, Huth & Ghumman, 2011; Barnes, Gunia & Wagner, 2015; Bereby-Meyer & Shalvi, 2015; Biziou-van-Pol et al., 2015; Cappelen et al., 2013; Capraro, 2017; Childs, 2012; Dreber & Johannesson, 2008; Erat & Gneezy, 2012; Friesen & Gangadharan, 2012; Gino, Schweitzer, Mead & Ariely, 2011; Gunia, Wang, Huang, Wang & Murnigham, 2012; Mead, Baumeister, Gino, Schweitzer, & Ariely, 2009; Shalvi, Eldar & Bereby-Meyer, 2012; Tatatabaeian, Dale & Duran, 2015; van't Veer, Stel & van Beest, 2014).

---

[1] See https://www.fbi.gov/stats-services/publications/insurance-fraud/insurance_fraud

Understanding which factors influence dishonest behavior is thus important for designing institutions to encourage honest behavior and discourage dishonest behavior. Here, I focus on the role of time pressure, which is a particularly relevant factor to be investigated in terms of both practical and theoretical applications. In practice, because people often have very little time to think through their decisions. This may happen both in social interactions, in which people have an incentive to decide quickly because thinking carefully about the available choices signals self-regarding motivations (Capraro & Kuilder, 2015; Hoffman, Yoeli & Nowak, 2015; Jordan, Hoffman, Nowak & Rand, 2016), and in economic interactions, in which acting fast may be crucial to overcome competitors. For example, traders are required to make decisions within seconds after new information is obtained (Busse & Green, 2002; Kocher, Pahlke & Trautmann, 2013; Roth & Ockenfels, 2002). In theory, because one recent framework (the Social Heuristics Hypothesis, SHH, Rand, Greene, & Nowak, 2012; Rand et al., 2014; Rand et al., 2016) makes clear predictions about what we should expect when forcing people to decide between honesty and dishonesty under time pressure vs time delay. The SHH argues that people internalize strategies that are optimal in their everyday interactions and tend to use them as default strategies in new and atypical situations when they have no time (or, more generally, no cognitive resources) to find out which choice maximizes their payoff. Then, after deliberation, people may override their heuristics and shift their behavior towards the one that is individually optimal in the given interaction. What does the SHH predict in terms of deceptive behavior in one-shot interactions? Of course, the optimal strategy in the given, one-shot interaction is to lie (in this paper, I focus on black lies, that is, lies that benefit the liar at the expenses of another person). Thus, the SHH predicts that deliberation favors deception. On the other hand, time pressure may prevent subjects from calculating their payoff-maximizing strategy. Thus, the SHH predicts that time pressure favors social heuristics that are optimal in everyday interactions. Since most daily interactions are repeated (e.g., with friends, family members, co-workers), truth-telling, although costly in the short term, may be optimal in the long run (through numerous channels, including the social stigma that accompanies liars). Thus, the SHH predicts that time pressure should favor truth-telling.

*Hypothesis.* Time pressure favors honesty in one-shot interactions.

In this paper, I present a large study in support of this hypothesis. To the best of my knowledge, this is the first study exploring this question. Two earlier studies have investigated the role of time pressure on honesty (Gunia et al., 2012; Shalvi et al., 2012); however, neither of them can be applied to our case, because participants in these experiments were communicated their payoff maximizing choice *before* the time manipulation[2]. Thus, time pressure did not limit

---

[2]In Shalvi et al. (2012) subjects were asked to report the outcome of a privately rolled die, knowing that they would be paid an amount of money equal to the reported outcome. The timer started after rolling the die. Thus, participants knew before the time manipulation that their payoff maximizing strategy was to report the number 6, regardless of the actual outcome of the dice. A conceptually similar design was implemented by Gunia et al. (2012), where

participants' ability to compute their payoff maximizing choice, which is the underlying requirement to apply the logic of the SHH.

**Measure of honesty**

To measure honest behavior, I use the Deception Game introduced by Biziou-van-Pol et al. (2015), which is a variant of the standard Deception Game (Gneezy, 2005; Erat and Gneezy, 2012). In this variant, participants are told that they *will* be randomly assigned to either Group 1 or Group 2, and that they *will* have to choose between two possible strategies: "telling the number of the group they are assigned to" or "telling the number of the other group". If they report the true number of the group they are assigned to, then both themselves and a randomly selected participant will get $0.10; otherwise they will get $0.20 and the other participant will get $0.09.

I have deliberately chosen to conduct the experiment with small stakes because previous research has shown that stakes have no effect on participants' behavior, as long as they are positive and not "too high". Specifically, it has been suggested that participants' behavior changes when passing from no-stakes to small stakes (Forsythe, Horowitz, Savin, & Sefton, 1994; Amir, Rand & Gal, 2012), then it is stake-independent at intermediate stakes, and then changes again when stakes approach one month of salary (Andersen, Ertac, Gneezy, Hoffman, & List, 2011; Kocher, Martinsson, & Visser, 2008), although the existence of the latter

---

participants were told that there were two available allocations of money, Option A and Option B; *senders* were informed that Option A would allocate $10 to themselves and $5 to the *receiver*, while Option B would allocate $5 to themselves and $10 to the receiver. Senders were then told they had to choose a message to send to the receiver, between "Option A earns you more money than Option B" and "Option B earns you more money than Option A". The role of the receiver was to guess which option would maximize their own payoff. After learning these pieces of information, senders moved to the decision screen, where some were asked to decide under time pressure and others were asked to decide under time delay. Also in this case, whatever their beliefs about the behavior of the receiver are, participants knew before the time manipulation their payoff maximizing strategy.

discontinuity is still under debate, since other studies have found that stakes do not matter even when they grow very large (Cameron, 1999).

**Method**

American subjects were recruited using the online platform Amazon Mechanical Turk (AMT). They earned $0.30 for completing the survey, plus an additional bonus depending on the choice they made in the Deception Game. Although AMT experiments are easy and cheap to implement and experimenters have much less control on participant's behavior during the experiment, several studies have shown that data gathered using AMT are of no less quality than those collected using the standard physical laboratory (Horton, Rand, Zeckhauser, 2011; Paolacci & Chandler, 2014).

After reading the instructions, all subjects faced the same set of comprehension questions. Subjects failing any comprehension question were automatically excluded from the survey. Subjects who passed the comprehension questions were randomly assigned to play a one-shot anonymous Deception Game either under *time pressure* condition or under *time delay*. Subjects under time pressure were asked to decide within 5 seconds; those under time delay were asked to stop and think for at least 30 seconds before deciding. Importantly, the number of the group a participant was assigned to was communicated directly in the decision screen. Thus, when the time manipulation started, participants knew that their optimal strategy was to deceive, but they did not know which choice corresponded to that strategy; in other words, time pressure worked as a limitation for participants' ability to compute their payoff maximizing choice. Decisions were collected using a blank text box in which subjects could to type their choice. Three sessions of the same study were conducted, one between Dec 15 and Dec 19, 2015, one between Feb 3 and Feb 8, 2016, and one on Nov 28, 2016. Each subject was allowed to participate in only one session. I refer the reader to the Appendix for full experimental instructions.

**Results**

A total of 1,013 participants (51.6% males, mean age = 35.36) passed the comprehension questions and participated in the experiment (N=497 under time pressure, N=516 under time delay). The time manipulation was successful, as subjects acting under time pressure took, on average, much shorter to make a decision than subjects under time delay (10.29s vs 31.78s, p<.001). As in previous studies (Tinghög et al., 2013; Rand, Newman & Wurzbacher, 2015; Capraro & Cococcioni, 2015), I include in the analysis also subjects who failed to obey the time constraints, in order to avoid selection problems[3]. Figure 1 provides visual evidence that subjects under time pressure were more honest than those under time delay (56.7% vs 44.2%). This is confirmed by logit regression predicting the probability of telling the truth as a function of a dummy variable, named "pressure", which takes value 1 if a subject acted under time pressure, and 0 otherwise ($chi^2$=16, coeff=0.505, p<.001). The positive effect of time pressure was essentially constant across sessions (Session 1: 57.8% vs 44.7%; Session 2: 56.4% vs 44.1%; Session 3: 55.9% vs 44.8%; all p's>0.5).

---

[3]Excluding from the analysis subjects who disobeyed the time constraints is highly inappropriate, because the reasons for which subjects fail to respect the time pressure may be different from those for which they fail to respect the time delay. In the time pressure case, subjects may fail to respect the time constraint simply because they did not have time to read the instructions; in the time delay condition, disobeying the time constraint is a deliberate choice. In other words, disobeying the time constraint in the time pressure condition may not have any effect on participants' behavior, while disobeying the time constraint in the time delay condition may make participants more similar to those in the time pressure condition. In support of this asymmetry of motivations for disobeying the time constraints, I indeed find that people who disobeyed the time constraint in the time pressure condition were as honest as those who obeyed it ($chi^2$=2.07, coeff=0.643, p=0.151), while those who disobeyed the time constraint in the time delay condition were significantly more honest than those who obeyed the time constraint ($chi^2$=6.24, coeff=0.623, p=0.013) and were as honest as those in the time pressure ($chi^2$=1.09, coeff=1.209, p=0.296).

Next, I observe that behavior under time pressure is in fact driven by truth-telling and not by confusion. If the effect of time pressure were driven only by confusion and not by honesty, then subjects under time pressure would choose randomly between honesty and dishonesty. But this is not the case: the proportion test shows that truth-telling under time pressure is significantly higher than 50% (56.7% +/- 2%, 95% CI = [52.4%,61.1%], p=0.001).

As an additional analysis, I also observe that when I include (log of) response time as dependent variable in the logit regression above (the log allows me to take into account the fact that response times follow a heavily right-skewed distribution), I find that time pressure retains significance (p=0.036), while response time is not significant (p=0.349). This is a somewhat interesting results, since recent research has highlighted that response times may be influenced by factors other than the extent of intuitive vs deliberative thinking, as, for example, decision conflict and strength of preferences (Krajbich, Bartling, Hare & Fehr, 2015; Evans, Dillon & Rand, 2015). Thus, the fact that the positive effect of time pressure is not mediated by response time provides another piece of evidence in support of the interpretation that time pressure favors *intuitive* honesty.

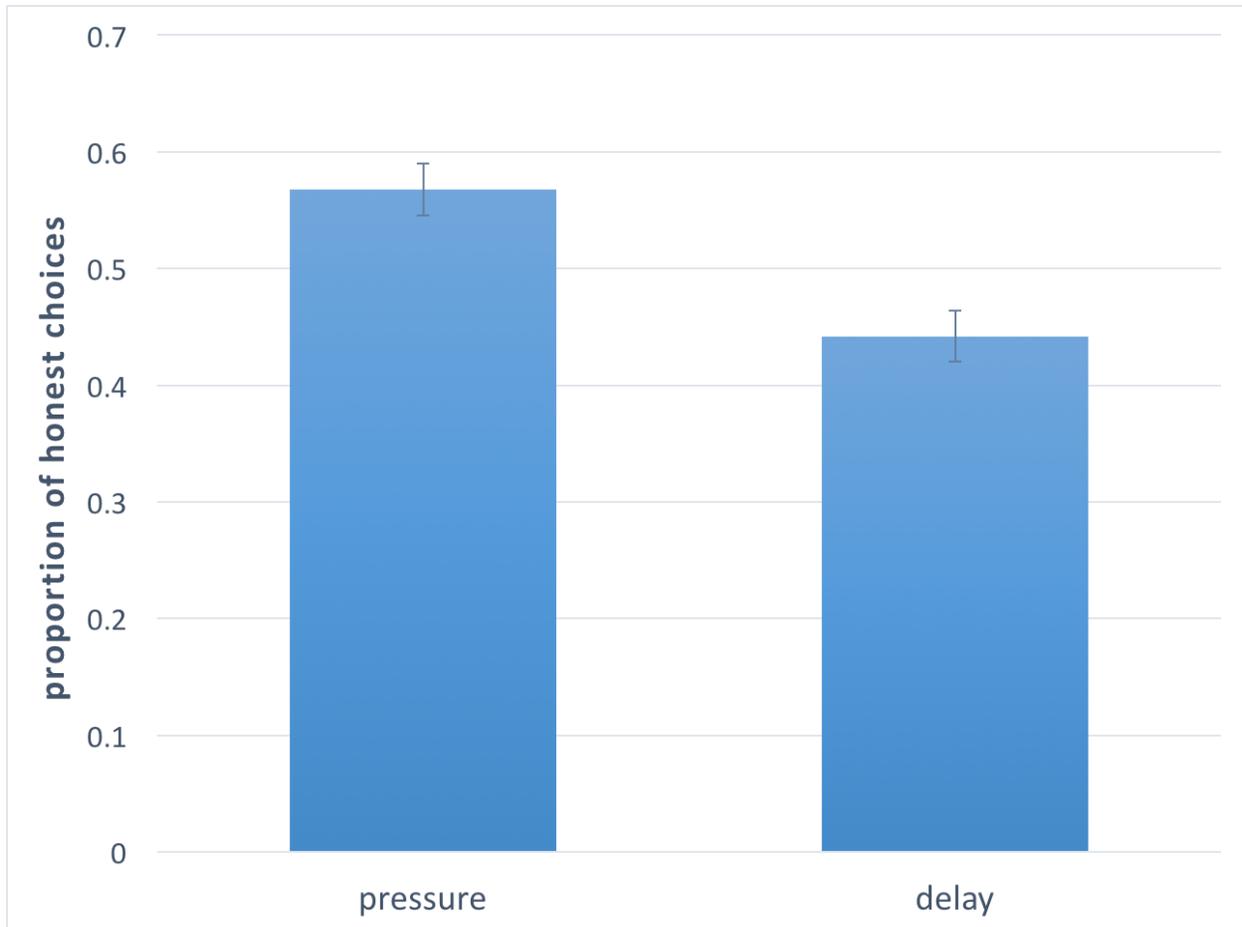

*Figure 1. Subjects under time pressure were more honest than those under time delay. Error bars represent +/- SEM.*

**Conclusion**

    Motivated by practical and theoretical considerations, I have explored whether time pressure affects honest behavior in one-shot deception games. Through a large study involving more than a thousand American subjects, I have demonstrated that this is indeed the case: time pressure makes people more honest. This result is in line with the Social Heuristics Hypothesis (Rand et al., 2014), a theory of social heuristics according to which people's default strategies correspond to choices that are optimal in their everyday interactions and that have been

internalized over time. Since most of our daily interactions are with friends, family, and coworkers, and thus they are repeated, being honest, although costly in the short run, may be optimal in the long run and get internalized as a default strategy.

References

……

**Appendix**

*Experimental instructions*

<u>Introductory screen (common to all conditions)</u>

Welcome to this HIT.

This HIT will take about two minutes. For the participation to this HIT, you will earn 40c. You can also earn additional money depending on the decisions that you will make.

To make sure you understand the situation, we will ask some simple questions, each of which has only one correct answer. If you fail to correctly answer any of those questions, the survey will automatically end and you will not receive any redemption code and consequently you will not get any payment.

With this in mind, do you wish to continue? (available answers: continue/end).

<u>Instruction screen (common to the time pressure condition and the time delay condition)</u>

Please read these instructions carefully. You may earn a considerable sum of money, depending on the decisions you make in the experiment.

You have been matched with another participant. Your bonus and that of the other participant depends only on your choice. The other participant does not play any active role.

There are two groups (group 1, group 2). You will be randomly assigned to one of them. You will be informed of the group you have been assigned to, but the other participant will not.

We will ask you to declare the group you have been assigned to. So you get to choose between two possible options:

Option 1: "I have been assigned to group 1."
Option 2: "I have been assigned to group 2."

If you choose the option corresponding to the number of the actual group you have been assigned to, then you will receive 10c and the other participant will receive 10c.

If you choose the option corresponding to the number of the other group (the one you have not been assigned to), then you will receive 20c and the other participant will receive 9c.

Finally, only you will be informed of the particular monetary value connected to each message. The other participant will not be informed of these monetary values.

*Comprehension questions (common to all conditions)*

1) What is the choice that maximize YOUR outcome? (available answers: Choosing the message corresponding to the number of the actual group you have been assigned to/Choosing the message corresponding to the number of the other group (the one you have not been assigned to).

2) What is the choice that maximize the OTHER PARTICIPANT'S outcome? (available answers: Choosing the message corresponding to the number of the actual group you have been assigned to/Choosing the message corresponding to the number of the other group (the one you have not been assigned to)).

*Decision screen for subjects under time pressure and assigned to group 1*

You have been assigned to group 1.

RESPOND WITHIN 5 SECONDS

Which group have you been assigned to?

(here there was a text box in which subjects could type their choice)

*Decision screen for subjects under time pressure and assigned to group 2*

You have been assigned to group 2.

RESPOND WITHIN 5 SECONDS

Which group have you been assigned to?

(here there was a text box in which subjects could type their choice)

*Decision screen for subjects under time delay and assigned to group 1*

You have been assigned to group 1.

THINK CAREFULLY FOR AT LEAST 30 SECONDS BEFORE CHOOSING

Which group have you been assigned to?

(here there was a text box in which subjects could type their choice)

*Decision screen for subjects under time delay and assigned to group 2*

You have been assigned to group 1.

THINK CAREFULLY FOR AT LEAST 30 SECONDS BEFORE CHOOSING

Which group have you been assigned to?

(here there was a text box in which subjects could type their choice).